\documentclass[12pt]{iopart}

\usepackage{graphicx}
\usepackage{caption}
\usepackage{cite}
\usepackage[colorlinks=true, letterpaper=true, pdfstartview=FitV, linkcolor=blue, citecolor=blue, urlcolor=blue]{hyperref}
\begin{document}

\title{Robust intrinsic half-metallic ferromagnetism in stable 2D single-layer MnAsS$_4$}

\author{Tengfei Hu, Wenhui Wan, Yanfeng Ge and Yong Liu}

\address{State Key Laboratory of Metastable Materials Science and Technology $\&$ Key Laboratory for Microstructural Material Physics of Hebei Province, School of Science, Yanshan University, Qinhuangdao 066004, China}
\ead{\mailto{yongliu@ysu.edu.cn}, \mailto{ycliu@ysu.edu.cn}}
\vspace{10pt}
\begin{indented}
\item[]February 2020
\end{indented}

\begin{abstract}
Two-dimensional (2D) intrinsic half-metallic materials are of great interest to explore the exciting physics and applications of nanoscale spintronic devices, but no such materials have been experimentally realized. Using first-principles calculations based on density-functional theory (DFT), we predicted that single-layer MnAsS$_4$ was a 2D intrinsic ferromagnetic (FM) half-metal. The half-metallic spin gap for single-layer MnAsS$_4$ is about 1.46 eV, and it has a large spin splitting of about 0.49 eV in the conduction band. Monte Carlo simulations predicted the Curie temperature (\emph{T}$_c$) was about 740 K. Moreover, Within the biaxial strain ranging from -5\% to 5\%, the FM half-metallic properties remain unchanged. Its ground-state with 100\% spin-polarization ratio at Fermi level may be a promising candidate material for 2D spintronic applications.
\end{abstract}
\noindent{\it Keywords\/}: First-principles calculations, 2D materials, half-metallic, spintronics
%\maketitle
\normalsize

\section{Introduction}

Spintronics, which uses the spin of electrons for the information storage, transport and processing, have attracted intensive interests from the viewpoint of fundamental science and technology applications in the past decades~\cite{1}. It is important in the field of quantum computing and the next-generation information technology~\cite{2,3}. Half-metallic materials, which is conducting in one spin orientation but insulating in the opposite spin direction meet the demand of a 100\% spin polarization ratio, are highly desirable for advanced spintronic applications~\cite{4}. The band gap for the insulating channel is termed as spin gap. To prevent spin leakage, the spin gap needs to be as wide as possible~\cite{5}. Since the first half-metallic material NiMnSb was predicted in 1983~\cite{6}, there has been a flurry of research into magnetic half-metals, such as transition metal compounds MnX (X = P, As), NbF$_3$, CoH$_2$, ScH$_2$, TiCl$_3$, VCl$_3$~\cite{7,8,9,10}; sp half-metallic ferromagnets RbSe and CsTe~\cite{11,12,13,14}.

Although half-metallic material has been studied for a long time, the demonstrated half-metals was very limited and have serious shortcomings, such as high cost or low \emph{T}$_c$. Until now, intrinsic half-metallic material with wide spin gap and high \emph{T}$_c$ is still absent in experiments. However, the single-layer CrPS$_4$~\cite{15} is predicted to be a ferromagnetic semiconductor and the valance bands are splited for different spin orientation. As hole doping can lower Fermi level into the valence bands of one spin and lead to half-metallic ferromagnets~\cite{16,17}. An obvious option is to replace Cr by Mn atoms, and by stability calculation we replace P with As atoms.

In this paper, the first-principles calculations are used to investigate the mechanical, dynamical, electronic and magnetic properties of single-layer MnAsS$_4$. Our calculations indicate that the MnAsS$_4$ crystal is mechanically and dynamically stable, so it is possibly prepared. It is metallic at the Fermi level in one spin direction and has a band gap of 1.46 eV in the opposite spin direction. Its half-metallic properties do not change under the biaxial strain range from -5\% to 5\%. It has stable ferromagnetism phase with integer magnetic moment of 8 $\mu_B$ per primitive cell. Furthermore, we demonstrated that the MnAsS$_4$ exhibits high \emph{T}$_c$ about 740 K.

\section{Methods}

Kohn-Sham DFT calculations are performed using the projector augmented wave method, as implemented in the plane-wave code VASP~\cite{18,19,20}. A cutoff energy of 500 eV and a Monkhorst-Pack special k-point mesh~\cite{21} of $17\times19\times1$ for the Brillouin zone integration was found to be sufficient to obtain the convergence. We used a Perdew-Burke-Ernzerhof (PBE) type generalized gradient approximation (GGA) in the exchange-correlation functional~\cite{22}. A conjugate-gradient algorithm was employed for geometry optimization using convergence criteria of 10$^{-7}$ eV for the total energy and 0.005 eV/{\AA} for Hellmann-Feynman force components. We used GGA+U to treat the strong on-site Coulomb interaction~\cite{23}. A series of U values were selected, that is, 1.0-6.0 eV for Mn. Whatever U was, it didn't affected the ground-state of MnAsS$_4$. Thus, whatever U were, the main conclusions were the same. So we displayed that results with Hubbard U term 5 eV for Mn~\cite{24} as suggested by Dudarev \emph{et al}~\cite{25}. The band structures for different U values can be found in Figure A1. Phonon dispersions were calculated by density functional perturbation theory~\cite{26} by the Phonopy package interfaced to VASP code with $2\times2\times1$ supercell. We inserted a 15 {\AA} vacuum slab to avoid the interactions between periodic images.

\section{Results and discussion}
\begin{figure}[htb]
  \centering
  \includegraphics[width=0.8\textwidth]{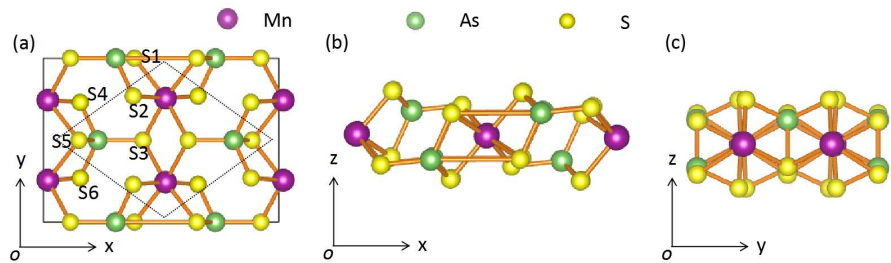}
 \caption{(Color online) The crystal structure of single-layer MnAsS$_4$ as seen from the (a) z-direction, (b) y-direction and (c) x-direction. The primitive cell is indicated by dotted line and the unit cell is indicated by solid line in (a).}\label{fig1}
\end{figure}

The atomic structure of single-layer MnAsS$_4$ is shown in Figure 1. The unit cell has parameters a = 11.34 {\AA}, b = 7.89 {\AA} (Figure C1) and space group C$_{2/m}$. As shown in Figure 1(b), the As atoms bridge the twisted MnS$_6$ octahedral chain, according to the crystal field theory, 3d orbitals of a metal ion in an octahedral ligand perturbation field split into t$_{2g}$ and e$_g$ orbitals, the ground electron configuration of Mn$^{3+}$ (t$_{2g}^3$ and e$_{g}^1$) leads to a half-metallic character. A Mn atom connects six S atoms, three types of bond lengths of Mn-S1, Mn-S2, Mn-S3 are 2.635\AA, 2.451{\AA} and 2.646\AA, respectively. An As atom connects four S atoms and the bond lengths of As-S3, As-S4, As-S5 are 2.198\AA, 2.182{\AA} and 2.214\AA, respectively, and the As-S6 bond length is the same as As-S4. In the diamond box in Figure 1(a) is a primitive cell. A primitive cell contains two Mn atoms. We calculated that each primitive cell is an integer magnetic moment of 8 $\mu_B$, and the local magnetic moment per Mn atom is about 4 $\mu_B$, it is consistent with previous work~\cite{27,28}

To determine the ground-state magnetic order, we compared the total energies of FM and different antiferromagnetic (AFM) structures~\cite{29}. The energy differences $\Delta$E relative to single-layer FM configurations are 170.69, 278.81, and 220.64 meV for the single-layer AFM1, AFM2, and AFM3 configurations, respectively. So the ground-state of single-layer MnAsS$_4$ is FM. Additionally, the non-magnetic (NM) state can be neglected owing to the great energy disparity between the NM state and magnetic states.
%The best place to locate the table environment is directly after its first reference in text
\begin{table}[h]%The best place to locate the table environment is directly after its first reference in text
 \newcommand{\tabincell}[2]{\begin{tabular}{@{}#1@{}}#2\end{tabular}}
    \centering
\caption{Elastic constants C$_{11}$, C$_{12}$ and C$_{22}$ (N/m) for single-layer MnAsS$_4$.}\label{table1}
\begin{tabular}{lccccc}
  \hline
   C$_{11}$ & C$_{12}$ & C$_{22}$\\
  \hline
   76.69 & 10.39 & 71.35\\
  \hline
\end{tabular}
\end{table}
\normalsize

Next, we determined its mechanical stability by calculating the three independent elastic constants. As shown in Table 1, we find that C$_{11}$ = 76.69 N/m, C$_{12}$ = 10.39 N/m and C$_{22}$ = 71.35 N/m, respectively. The elastic constants clearly satisfy Born¡¯s stability criterion~\cite{30}, i.e., C$_{11}{>}$0, C$_{22}{>}$0 and C$_{11}$-C$_{12}{>}$0, indicating that they are mechanically stable.
Meanwhile, we evaluate the stability of single-layer MnAsS$_4$ by comparing their binding energies,  which is defined as
\[{E_b} = \frac{{2E(Mn) + 2E(As) + 8E(S) - E(MnAs{S_4})}}{{12}},\] where \emph{E}(\emph{Mn}), \emph{E}(\emph{As}), \emph{E}(\emph{S}) and \emph{E}(\emph{MnAsS$_4$}) are the energy of Mn atom, As atom, S atom and single-layer MnAsS$_4$, respectively. According to this theory, the bigger \emph{E}$_b$ is, the more stable the system will be. We find the binding energy is 4.03 eV per atom, which is bigger than the synthetic VI$_3$~\cite{28} and others~\cite{31,32}.

In order to ensure the single-layer MnAsS$_4$ is dynamically stable, we calculated its phonon dispersion. As shown in Figure B1, the imaginary frequency is found to be absent in the whole Brillouin zone, it suggests that single-layer MnAsS$_4$ is dynamically stable and can exist as free-standing 2D crystal.

\begin{figure}[htb]
\centering
\includegraphics[width=0.8\textwidth]{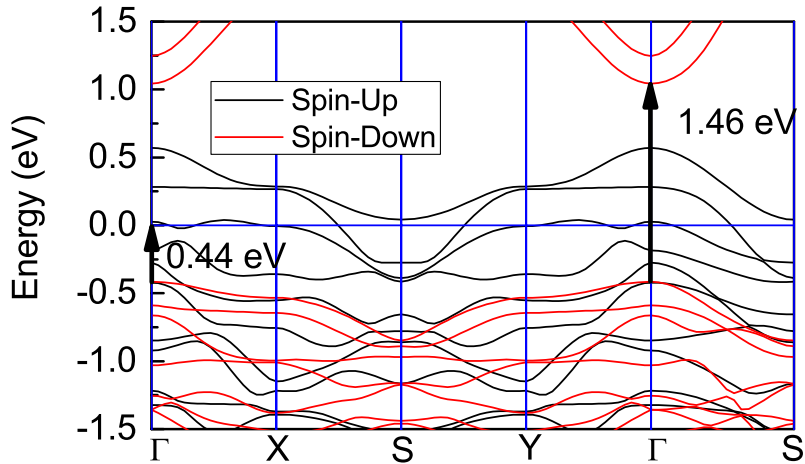}
\caption{(Color online) Electronic band structures for single-layer MnAsS$_4$.}\label{fig2}
\end{figure}
Next, the electronic properties of single-layer MnAsS$_4$ were investigated. The band structures of single-layer MnAsS$_4$ are shown in Figure 2. Notably, the spin-up bands cross the Fermi level, while the spin-down channel acts as a semiconductor, indicating that it is intrinsic half-metallic material with 100\% spin-polarization ratio. Comparing with previous studies where half-metallic materials occurred under certain external constraints, the half-metallic material found here is totally intrinsic, meaning that single-layer MnAsS$_4$ is more suitable for actual spintronic applications. As mentioned before, a wide half-metallic band gap and a  wide spin gap are very important for half-metal in spintronic applications~\cite{33,34}. Herein, the half-metallic band gap for single-layer MnAsS$_4$ is 0.44 eV, which is larger and smaller than the previous research on TiCl$_3$ (0.42 eV) and on VCl$_3$ (0.64 eV)~\cite{10}. The spin gap for the semiconducting channel is 1.46 eV, which is larger than the previous report of Fe$_2$Si~\cite{35}. The wide spin gap and half-metallic gap make 2D MnAsS$_4$ an ideal candidate for miniaturized spintronic materials. Considering that, approximately 25\%-45\% of the spin gap is underestimated by DFT method~\cite{36}, the spin gap of single-layer MnAsS$_4$ obtained by experiment may be larger than the above value. However, the trends of band dispersions and density of states are qualitatively reasonable, since we are mainly interested in their relative values, and the underestimate should not change the general trends of the results~\cite{37,38}.

\begin{figure}
  \centering
  \includegraphics[width=0.8\textwidth]{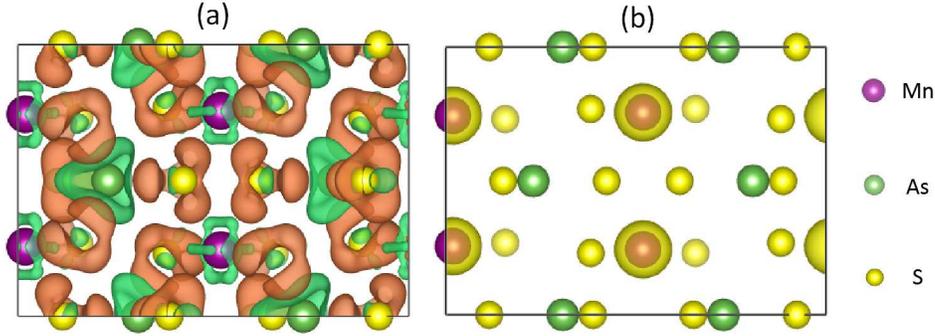}
 \caption{(Color online) (a) Charge density difference isosurface plot, showing charge accumulation and depletion (brown and green respectively). (b) Spin density isosurface plots for FM spin arrangements. }\label{fig3}
\end{figure}

Figure 3(a) shows the charge density difference of single-layer MnAsS$_4$. It is the difference between the charge density at the bonding point and the atomic charge density at the corresponding point. The brown and green region represent the charge accumulation and depletion. It is obvious that Mn and As atoms lose electrons and S atoms gain electrons, due to that S atom has larger electronegative. This allows Mn-S bonding to be more ionic. Figure 3(b) shows the spin density of single-layer MnAsS$_4$, we find that the induced spin polarization is mainly contributed by Mn atoms while the contribution from As and S atoms can be neglected, which is consistent with the magnetic moment analysis.
\begin{figure}[htb]
    \centering
  \includegraphics[width=0.8\textwidth]{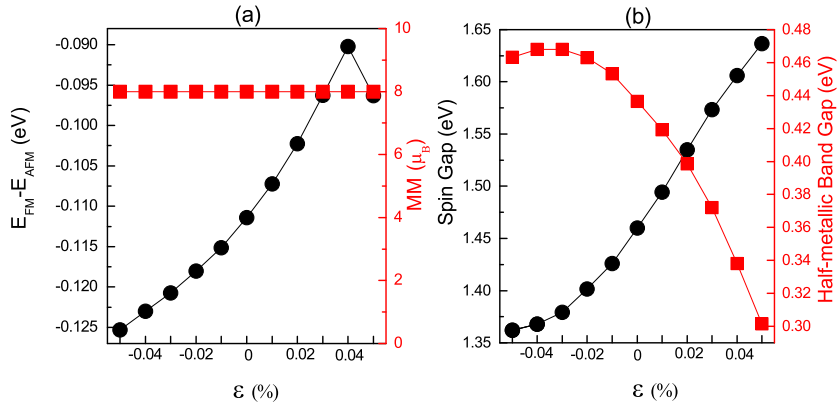}
    \caption{(Color online) Under biaxial strain for single-layer MnAsS$_4$. (a) Energy difference between the FM and AFM phases (black line) and total magnetic moments (red line). (b) Variation of the spin gap (black line) and half-metallic band gap (red line).}\label{fig4}
  \end{figure}

Figure 4(a) (red line) shows the change of total magnetic moment when a biaxial strain is applied. We find that the total magnetic moment is not affected when the biaxial strain range from -5\% to 5\%. Figure 4(a) (black line) shows the energy difference between FM and AFM configurations, and we can see that all values are negative, which indicates that no phase transition occurs during the process of biaxial strain and FM is always ground-state. As shown in Figure 4(b) (black line), when the biaxial stretch occurs, the spin gap increases, but decreases when the biaxial compression. When the strain is 5\%, the spin gap increases from 1.46 eV to 1.64 eV, and when the strain is -5\%, the spin gap decreases to 1.36 eV, this trend is consistent with the previous report of CrSiTe$_3$~\cite{39}. When the biaxial strain of -3\% is applied, the half-metallic band gap reaches the maximum value of 0.47 eV. When the biaxial strain is 5\%, the half-metallic band gap is 0.30 eV.

\begin{figure}[htp]
  \centering
  \includegraphics[width=0.8\textwidth]{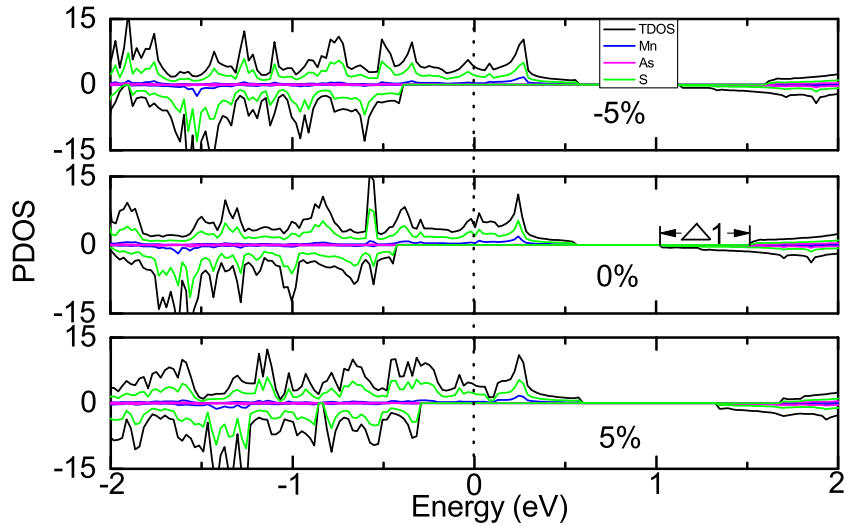}
 \caption{(Color online) Density of states for single-layer MnAsS$_4$ under different biaxial strain. The percentage in the figure indicates the magnitude of the strain.}\label{fig5}
\end{figure}
 For the single-layer MnAsS$_4$, a large spin exchange splitting of 0.49 eV (labeled as $\Delta1$ in Figure 5) in the conduction band is observed, which is crucial for the application in spin-polarized carrier injection and detection~\cite{40}. When applying a biaxial strain of -5\% and 5\%, spin exchange splitting changes from 0.49 eV to 0.47 eV and 0.35 eV, respectively, which is larger than the CrGeTe$_3$ (0.24 eV). Whether it is compressed or stretched, the density of states clearly shows that the contributions near the Fermi level are mainly from the Mn and S atoms.

 \begin{figure}[htp]
  \centering
  \includegraphics[width=1\textwidth]{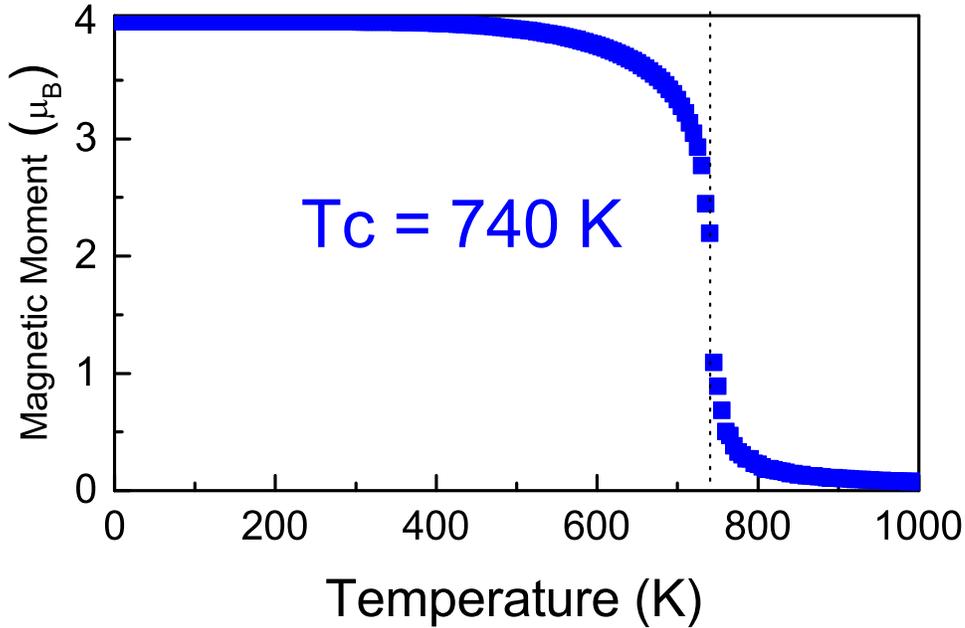}
 \caption{(Color online) Monte Carlo simulations on the average magnetic moment of Mn atoms for single-layer MnAsS$_4$ based on Ising model.}\label{fig6}
\end{figure}
Now, we turn our attention to the microscopic origin of the exchange interactions. Along the y direction, the distance between Mn atoms is too large (4.044{\AA}) for the direct-exchange coupling to play an important role, the optimized Mn-S-Mn angle is 99.6$^\circ$, which is near the ideal 90$^\circ$ bond angle. 90$^\circ$ bond angle usually associated with FM ordering and 180$^\circ$ associated with AFM ordering, according to the Goodenough-Kanamori rules~\cite{41,42}, so it exhibits FM properties. Along the x direction, the exchange interactions are more indirect because the Mn-Mn distance is even larger (5.668{\AA}). This indirect-exchange coupling is not only mediated by S atoms but also by As atoms, the optimized Mn-S-As, S-As-S, As-S-Mn angle is 110.7$^\circ$, 120.2$^\circ$, 83.4$^\circ$, respectively. The transition crossover angle of FM-AFM is 127$\pm$0.5$^\circ$~\cite{43}, and the bond angle along the x direction are all smaller than 127$^\circ$, so they prefer FM order. Therefore, the system presents the FM ground-state.

To realize spintronic applications for singer-layer MnAsS$_4$, it is necessary to obtain the variation trend of local magnetic moment with \emph{T}$_c$. We thus employed Monte Carlo simulations with the Hamiltonian ${H =  - \sum\limits_{ < ij > } {{J_{ij}}{S_i}{S_j}}}$ to predicted the FM transition temperature, where $\emph{J}_{ij}$ represents the exchange interactions of over all neighbor Mn-Mn pairs, $\emph{S}_i$ represents the spin of atom $\emph{i}$. Herein we only consider nearest neighbors located along x and y directions. The exchange interaction parameters $\emph{J}_{ij}$ are determined by the relation between the total energy and spin configurations. Our result of total $\emph{J}_{ij}$ is 5.33 meV and 8.71 meV along the x and y directions, respectively, S=2. As shown in Figure 6, the \emph{T}$_c$ extracted from the figure is around 740 K. It is significantly higher than those reported before, e.g., CrI$_3$ monolayer (45 K)~\cite{44}, CrSiTe$_3$ (35.7 K) and CrGeTe$_3$ (57.2 K)~\cite{40}

In similar structures, the AFM ground-state causes the energy band of spin-up and spin-down to coincide, which may lead to the formation of semiconductor materials. We predicted that the ground-state of MnSbS$_4$, VPS$_4$ and CrAsS$_4$ is AFM, detailed studies focusing on the aspect might reveal some interesting physics.

%\begin{figure}
%\includegraphics{1.eps} % Here is how to import EPS art
%\caption{\label{fig1} A figure caption. The figure captions are
%aut
%omatically numbered.}
%\end{figure}

\section{Conclusions}

In summary, we presented a intrinsic ferromagnetic half-metallic material by using first-principles calculations. The calculations of mechanical properties, phonon dispersion and binding energy ensure the stability and the possibility of preparation of single-layer MnAsS$_4$. The band structures show that the single-layer MnAsS$_4$ has a 100\% spin-polarization ratio at Fermi level. Also, for the semiconducting channel, the spin gap and half-metallic band gap are 1.46 eV and 0.44 eV, respectively. It also has a large spin exchange splitting of 0.49 eV in the conduction band. Within the biaxial strain range from -5\% to 5\%, the ferromagnetic half-metallic properties remain unchanged. Monte Carlo simulations estimate that \emph{T}$_c$ for single-layer MnAsS$_4$ can up to 740 K. The intrinsic half-metallic with high \emph{T}$_c$ and excellent stability endows single-layer MnAsS$_4$ a promising functional material for spintronic applications.

\section{Acknowledgments}

%\section{\label{sec:level1}RESULTS AND DISCUSSION}
%\subsection{\label{sec:level2}Crystal structure and lattice dynamics}

This work was supported by National Natural Science Foundation of China (No.11904312 and 11904313),the Project of Hebei Educational Department, China(No.ZD2018015 and QN2018012), the Advanced Postdoctoral Programs of Hebei Province (No.B2017003004) and the Natural Science Foundation of Hebei Province (No. A2019203507).  Thanks to the High Performance Computing Center of Yanshan University.

\clearpage
\appendix
\appendix\renewcommand{\thefigure}{A\arabic{figure}}
\begin{figure}[htb]
  \centering
  \includegraphics[width=0.7\textwidth]{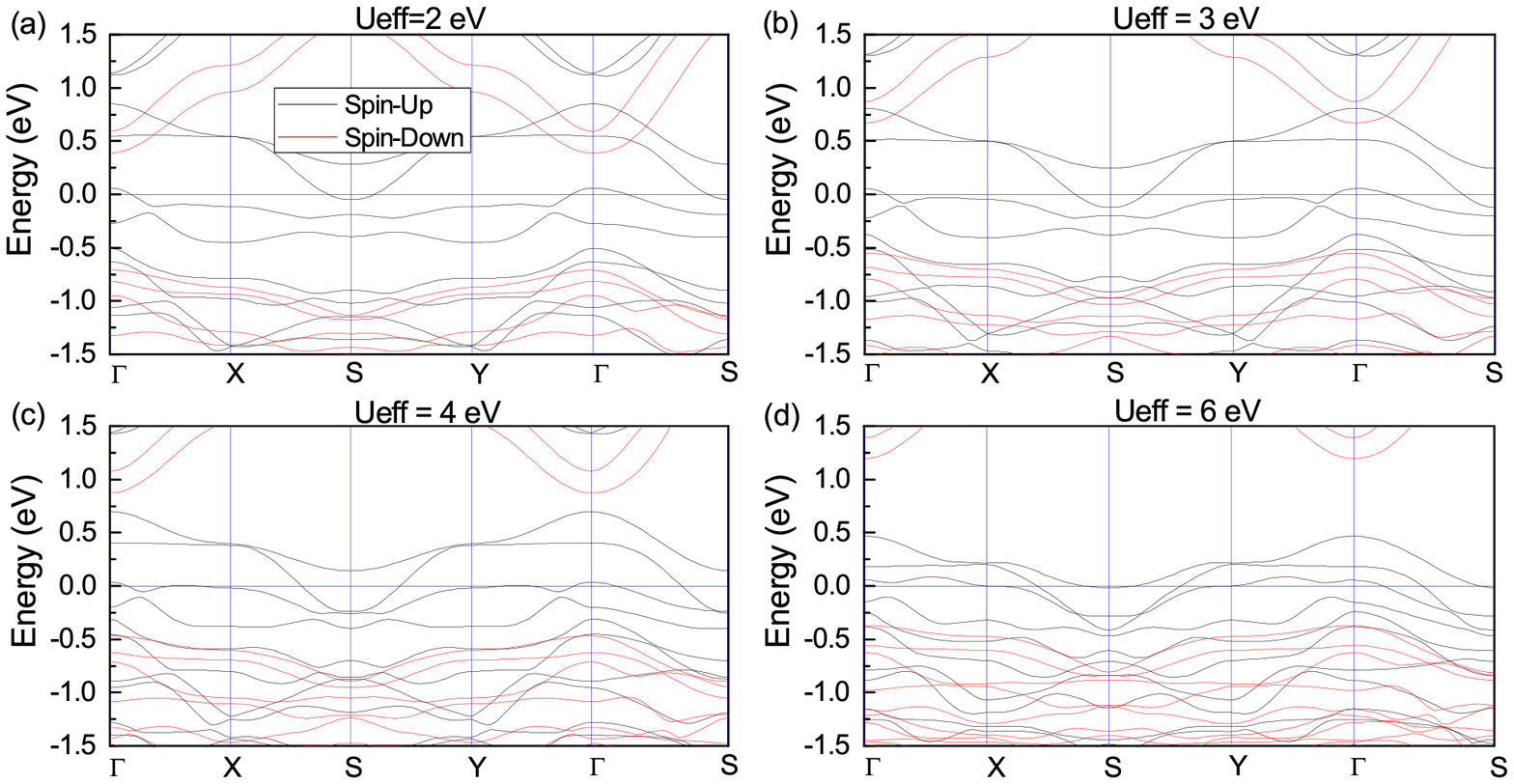}
 \caption{(Color online) The band structures of single-layer MnAsS$_4$ for different U values}\label{figA1}
\end{figure}
\appendix\renewcommand{\thefigure}{B\arabic{figure}}
\begin{figure}[htb]
  \centering
  \includegraphics[width=0.5\textwidth]{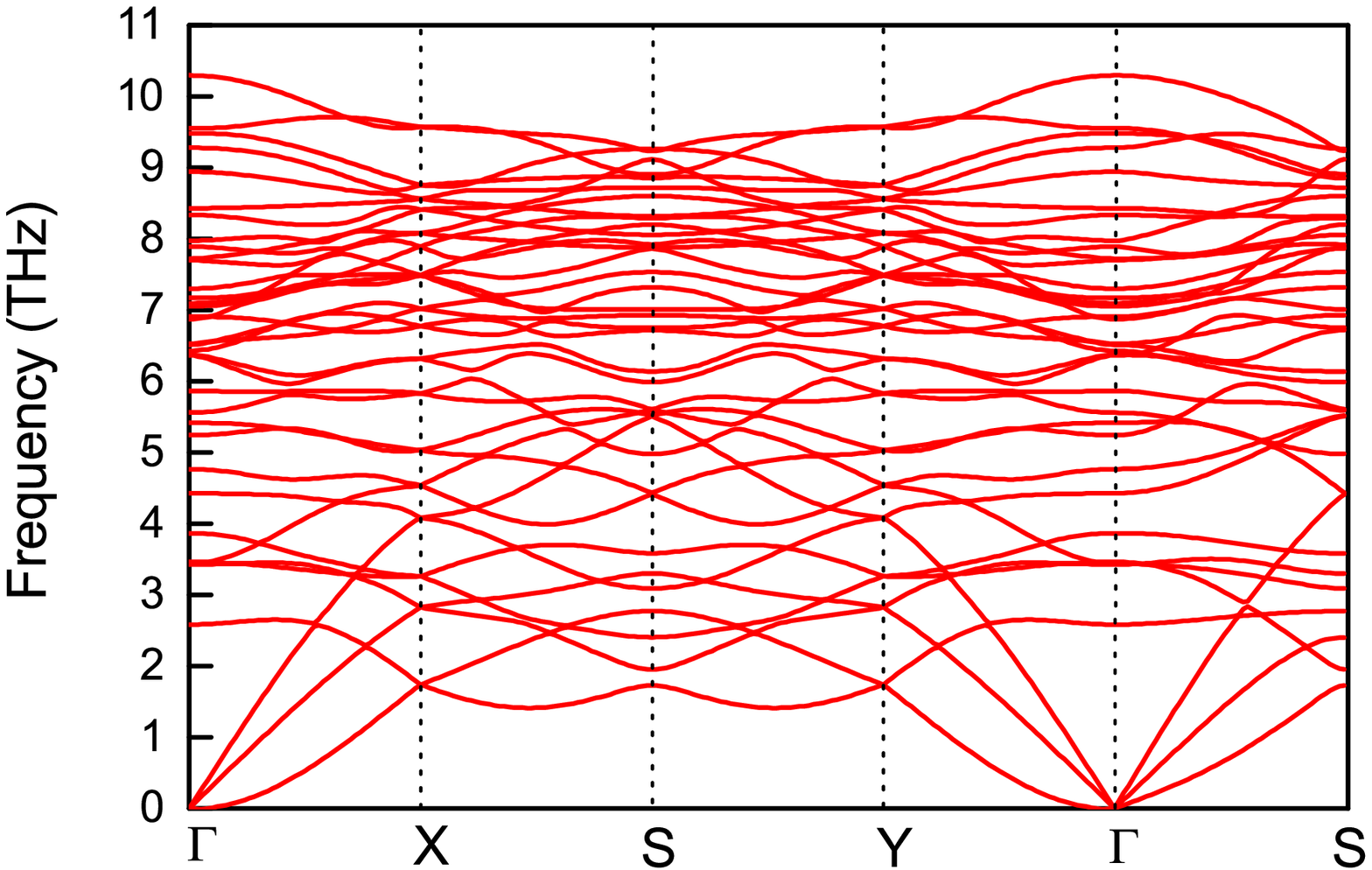}
 \caption{(Color online) Phonon dispersion of single-layer MnAsS$_4$ obtained from DFT calculations with the PBE functional.}\label{figB1}
\end{figure}
\appendix\renewcommand{\thefigure}{C\arabic{figure}}
\begin{figure}[htb]
  \centering
  \includegraphics[width=0.5\textwidth]{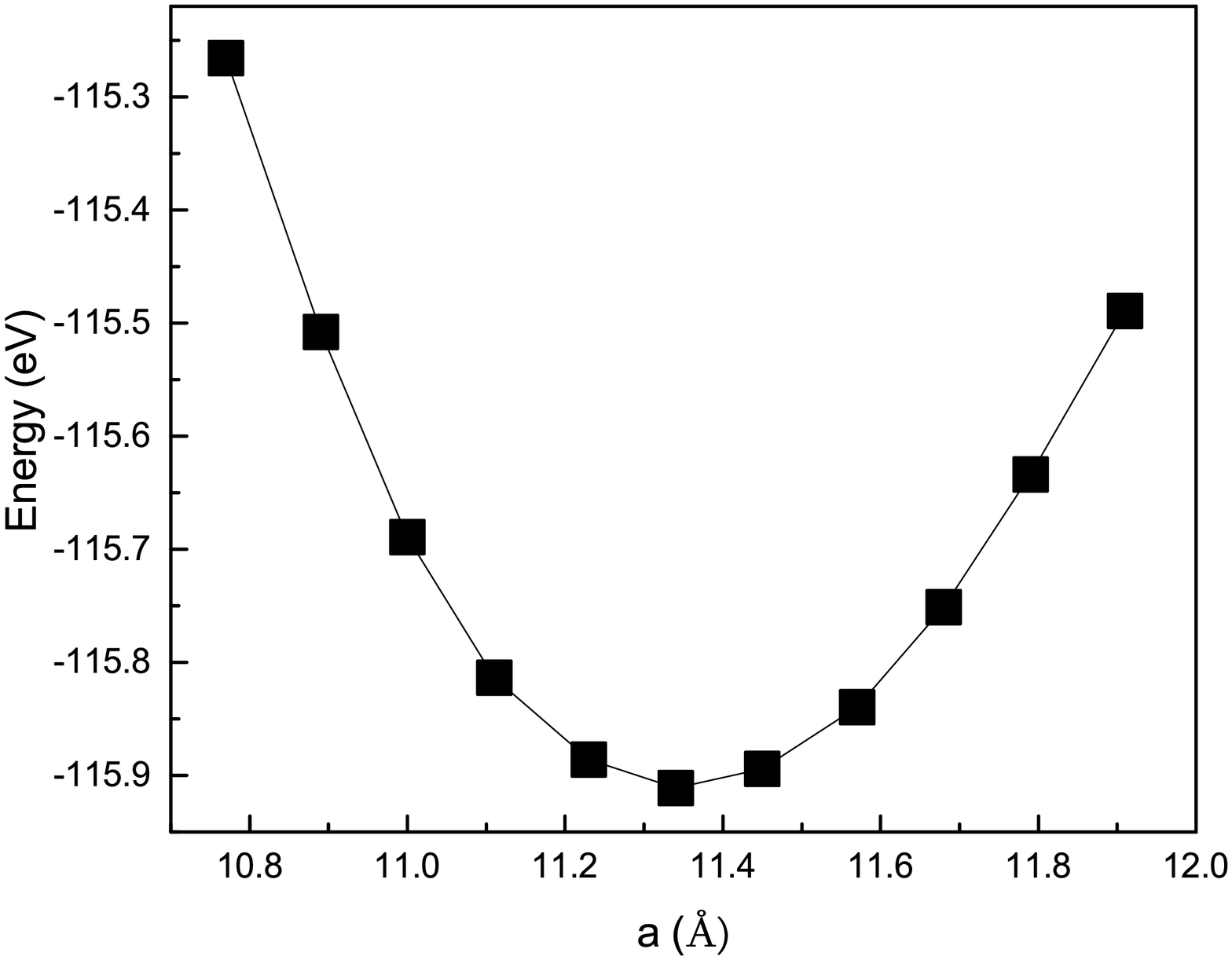}
 \caption{(Color online)  Variation of total energy with the lattice constant for single-layer MnAsS$_4$.}\label{figC1}
\end{figure}

%\indent\textcolor{blue}{\em Acknowledgments.}--


\section*{References}


\nocite{*}
\bibliographystyle{unsrt}

\bibliography{hu}

\end{document}